\documentclass[a4paper,11pt]{article}
\usepackage{pos}
\usepackage{graphicx}
\usepackage{graphics}
\usepackage[mathscr]{eucal}
\usepackage{amsbsy}
\usepackage{pifont}
\usepackage{amsmath}
\usepackage{xspace}
\usepackage{listings}
\usepackage{ulem}
\usepackage{slashed}
\usepackage{verbatim}
\usepackage{axodraw4j}
\usepackage{caption}
\usepackage{subcaption}
\usepackage{multirow}
\usepackage{xcolor}
\usepackage{hyperref}
\usepackage[capitalise]{cleveref}
\usepackage[title,toc]{appendix}
\usepackage{cancel}
\usepackage{soul}
\usepackage{natbib}
\usepackage{mciteplus}
\usepackage{lineno}
\usepackage{placeins}
\usepackage{comment}

\usepackage[shortlabels]{enumitem}

\definecolor{ultramarine}{rgb}{0.5, 0.125, 0.376}
\definecolor{mycolor}{rgb}{0.9, 0.5, 0.}

\input{Defs.sty}

\title{Scalar sector of Type 2 Seesaw model
explorations with multi-lepton final states}
\author[a]{C\u alin Alexa}
\author[a]{Otilia A. Ducu}
\author[a]{Ana E. Dumitriu}
\author*[a]{Adam Jinaru}
\author[b]{Emmanuel~Monnier}
\author[c]{Gilbert Moultaka} 
\author[a]{Alexandra Tudorache}
\author[d]{Hanlin Xu}

\affiliation[a]{IFIN-HH Bucharest; Romania}
\affiliation[b]{CPPM, Aix-Marseille Universit\'e, CNRS/IN2P3, Marseille; France}
\affiliation[c]{Laboratoire Univers \& Particules de Montpellier (LUPM), Universit\'e de Montpellier, CNRS, Montpellier; France}
\affiliation[d]{Department of Modern Physics and State Key Laboratory of Particle Detection and Electronics,
University of Science and Technology of China, Hefei; China}

\emailAdd{adam.jinaru@cern.ch}
\emailAdd{calin.alexa@cern.ch}
\emailAdd{otilia.anamaria.ducu@cern.ch}
\emailAdd{ana.dumitriu@cern.ch}
\emailAdd{monnier@in2p3.fr}
\emailAdd{gilbert.moultaka@umontpellier.fr}
\emailAdd{alexandra.tudorache@cern.ch}
\emailAdd{hanlin.xu@cern.ch}

\abstract{
   Originally motivated for the generation of (Majorana) neutrino masses, the Type 2 Seesaw Model has also a rich extended Higgs sector with, if accessible at the LHC, a distinctive phenomenology of neutral, charged and doubly-charged states. The goal of the work is to present an exhaustive phenomenological study of the most promising production and decay channels of pair or associated scalars, decaying directly or in cascades to Standard Model particles at the LHC. The study is complementary to the literature in that it highlights a previously unnoticed important sensitivity to a mixing angle. The ensuing uncertainty calls for a comprehensive experimental search strategy for the various processes. These  processes can be studied within LHC energies reach, by comparing cutflow results for different final states. We carried out prospective search analyses with multi-lepton, jets and missing energy configurations, assuming an ATLAS-like detector at LHC and HL-LHC, for charged, doubly-charged, and for the first time neutral scalar productions. The work is a collaboration between ATLAS experimentalists and theoreticians in continuation of an endeavor that lead to previous published ATLAS analyses for the search of (doubly)charged Higgs bosons~\cite{ATLAS2019,ATLAS2021}, aiming at proposals for future experimental searches. This proceeding only summarizes some of the results, and the reader is encouraged to consult the extended work \cite{ducu} for further detailing the content of each section.  
}

\FullConference{42nd International Conference on High Energy Physics (ICHEP2024)\\
18-24 July 2024\\
Prague, Czech Republic\\}

\begin{document}
\maketitle

\section{Introduction}

\par
Extending the SM Higgs sector by another Y=2 Higgs complex triplet allows a seesaw mechanism to account for small neutrino Majorana masses with order one Yukawa couplings. This is the so-called Type 2 Seesaw model(see \cite{ducu} and references therein). 
Typically, the mass scale of the triplet would be GUT-like, in which case the extra Higgs bosons would be out of reach at the colliders. However, much lower scales can be considered where the new scalars become reachable at the LHC, at the price of tuning down the neutrino Yukawa couplings. 
Type 2 Seesaw model has been extensively studied in the literature  (see e.g. \cite{ashanujjaman} and references therein).
It accounts naturally for a SM-like Higgs boson $h^0$ and 6 other heavier states: $H^0$, $A^0$, $H^\pm$, $H^{\pm \pm}$. An important parameter is the vacuum expectation value (VEV) of the triplet ($v_t$):  For illustration, we take $v_t=0.1$~GeV so that  lepton-number-violating decays of the scalars (e.g. decays to pairs of same charge leptons for which experimental limits are strong)  are suppressed. In this case, a wealth of  decays to vector bosons, scalars and possibly heavy quarks opens up.

\par
In the following we summarize the main results of the detailed study carried out in~\cite{ducu}. 
%In this proceeding, we  \gilbert{summarize} studies already done in~\cite{Marseille:2024ddj}: first scan on some of the parameters of the model so to calculate the scalars' widths and BRs expressions, then find out what the most promising BRs look like in order to combine them in such a way to get the most promising intermediate and final states, and finally do some  prospect studies on some chosen multilepton final states.

\section{The scanning procedure, production and decay modes of the scalars }
\label{sec:production_decay}

In order to study the various production cross-sections and decay branching ratios, one needs to scan over the masses of the scalars and their couplings. Apart from the known gauge and Yukawa couplings, a natural way of proceeding can be by scanning over the parameters of the extended Higgs potential to get the physical masses, mixing angles and couplings. Alternatively one could scan directly over the physical scalar masses and get the parameters of the potential. However, due to problems for both methods, a hybrid strategy was adopted, such that advantages of both methods are kept, and disadvantages discarded. In particular it allows to control directly the SM-like state mass, as well as its tree-level couplings from
the parameter \sina, where $\alpha$ denotes a mixing angle in the CP-even scalar sector, taken as input.

A scan of the production cross-sections at the LHC $\sqrt{s}=13$~TeV, for $H^{\pm \pm}$ in the 200 - 800 GeV mass range, shows the dominance of Drell-Yan pair or associated production modes. The associated $H^{\pm \pm} H^{\mp} $ production mode has the leading contribution,  decreasing with increasing $H^{\pm \pm}$ mass, being closely followed by the $H^0A^0$ and $H^{\pm \pm} H^{\mp \mp}$ productions. $H^\pm H^0 $  is the next most important channel, on equal footing with $H^\pm A^0$, since $H^0$ and $A^0$ are mass degenerate. However, this hierarchy can somewhat change by  varying the $|m_{H^{\pm \mp}} - m_{H^\pm}|$ mass splitting. The single scalar/$Z/W$ associated production modes are  suppressed by small $({v_t}/{v_d})^2$, where $v_d$ denotes the doublet VEV.

\begin{figure}[!ht]
    \begin{center}
      {\includegraphics[width=0.45\textwidth]{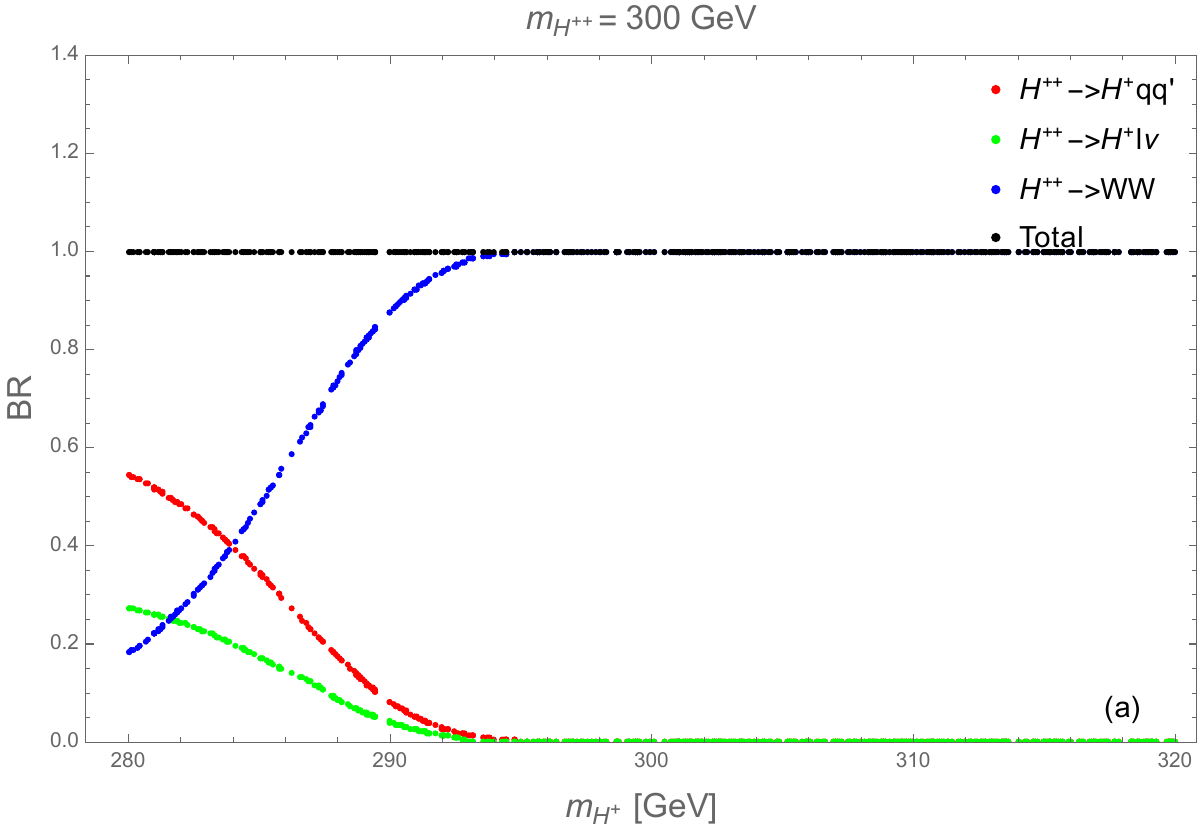}}  
       {\includegraphics[width=0.45\textwidth]{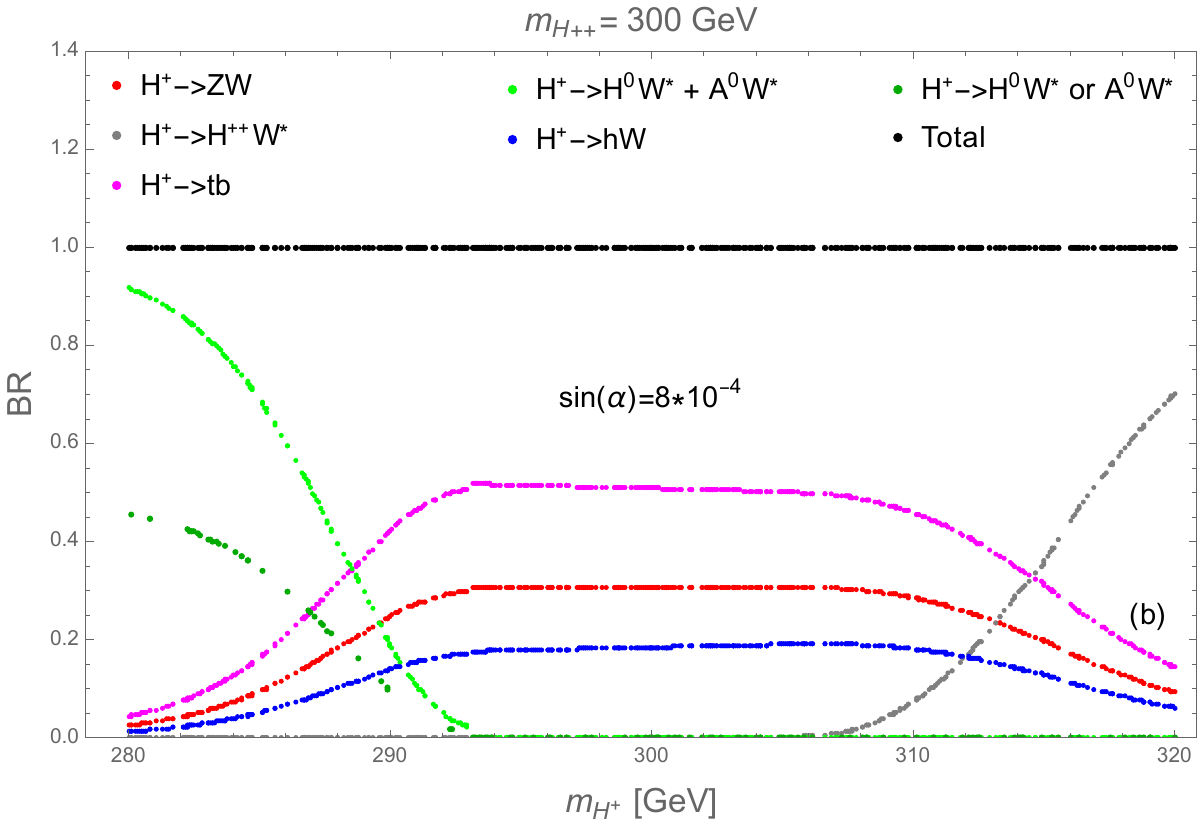}}  \\
      {\includegraphics[width=0.45\textwidth]{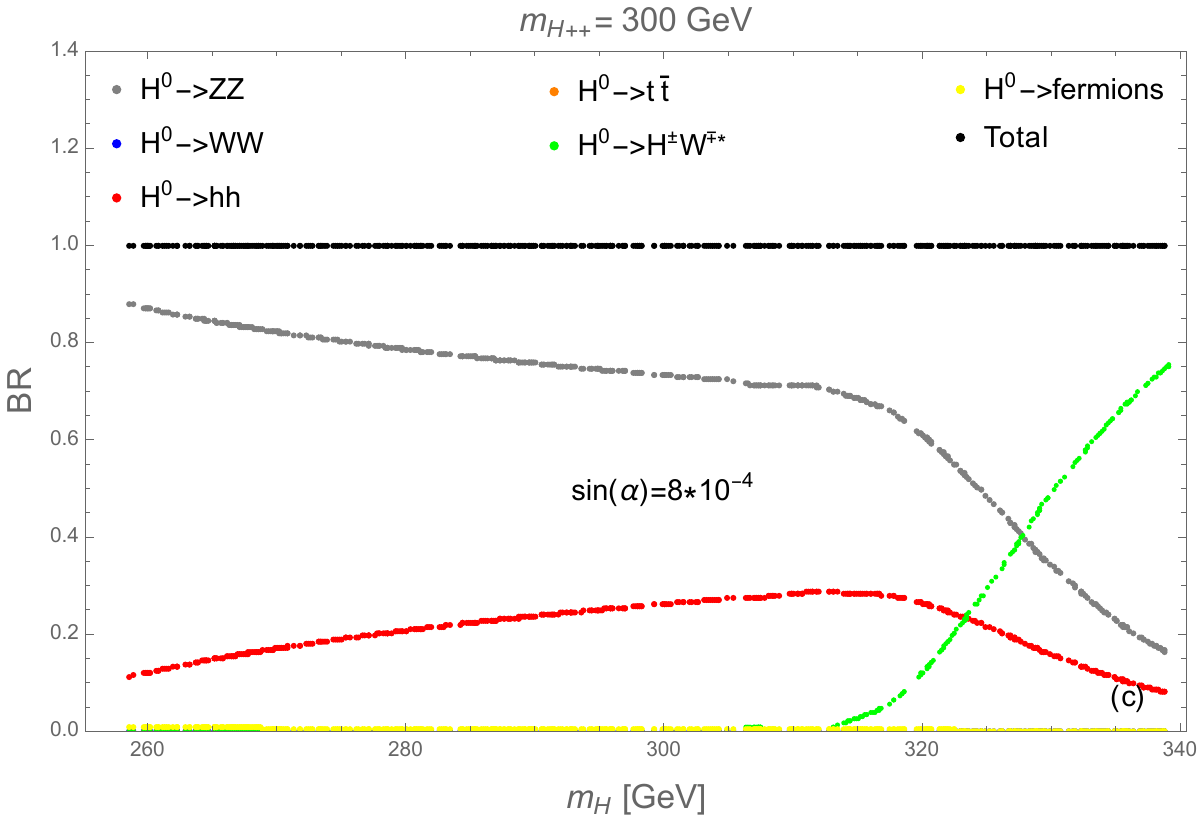}}   {\includegraphics[width=0.45\textwidth]{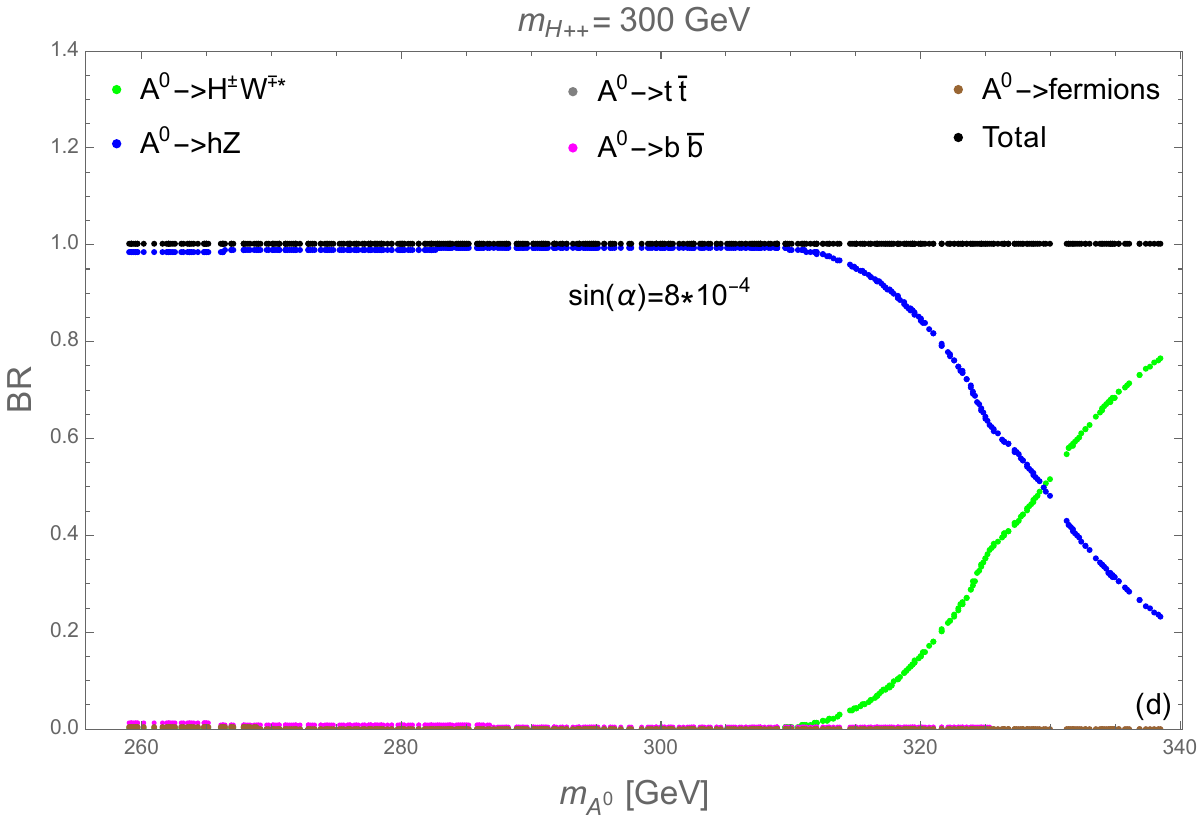}}
    \end{center}
\caption{\label{fig:BRs} Branching ratios for different Higgs scalars of the model, for \mHpp $=300$~GeV, $v_t=0.1$~GeV and \sina $=8 \times 10^{-4}$.}
\end{figure}

\begin{figure}[!bht]
    \begin{center}
      {\includegraphics[width=0.45\textwidth]{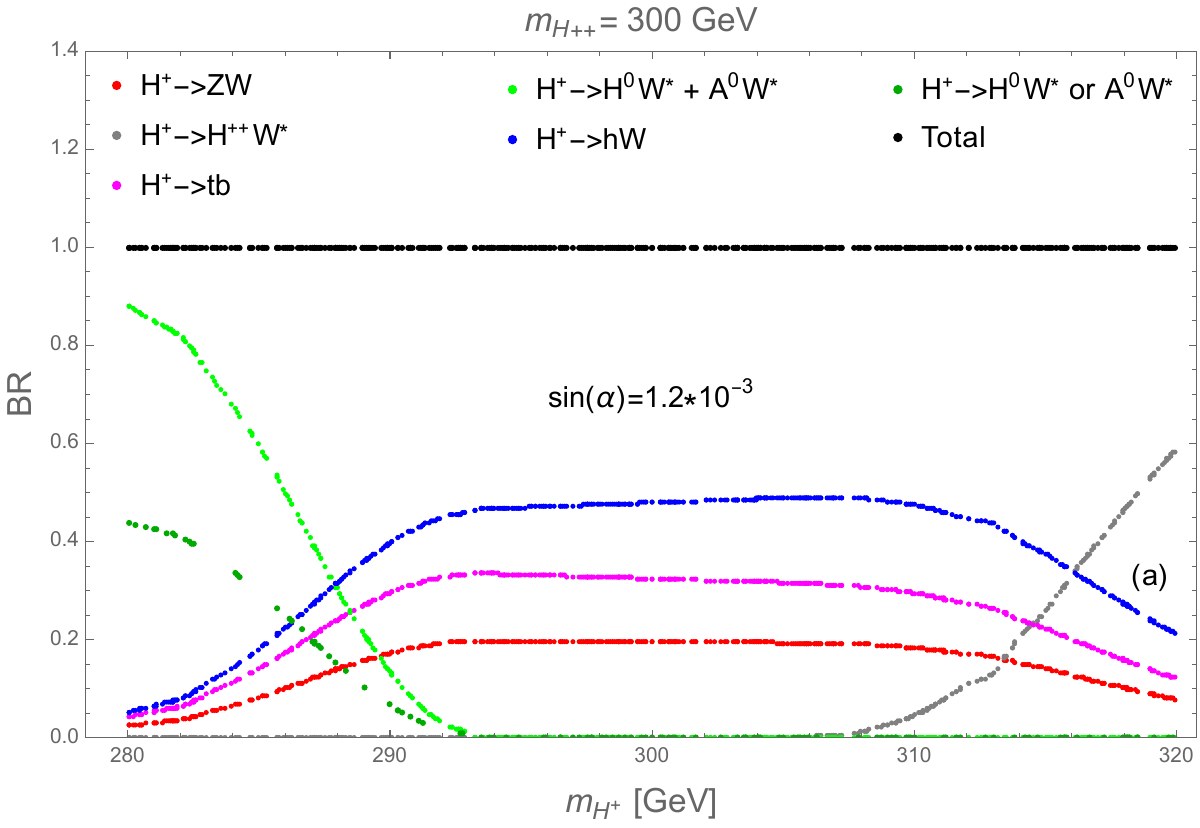}}
      {\includegraphics[width=0.45\textwidth]{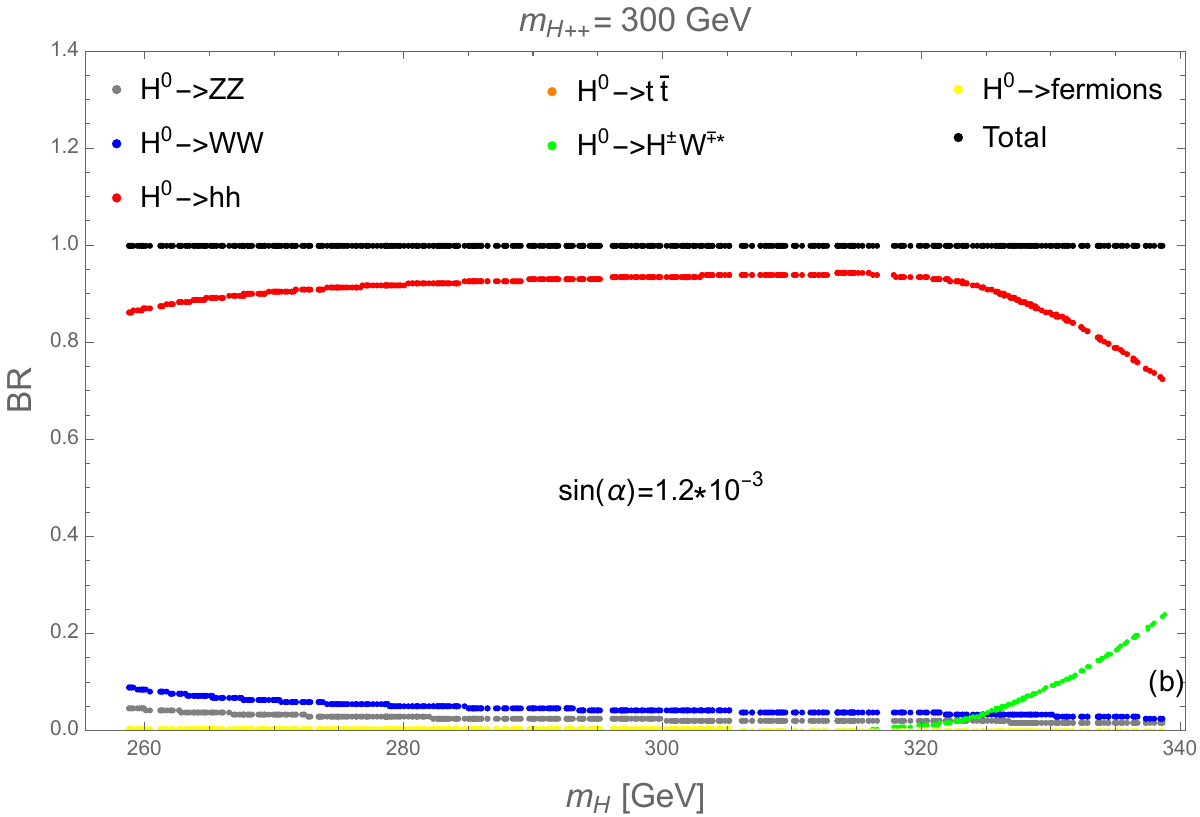}}\\   {\includegraphics[width=0.45\textwidth]{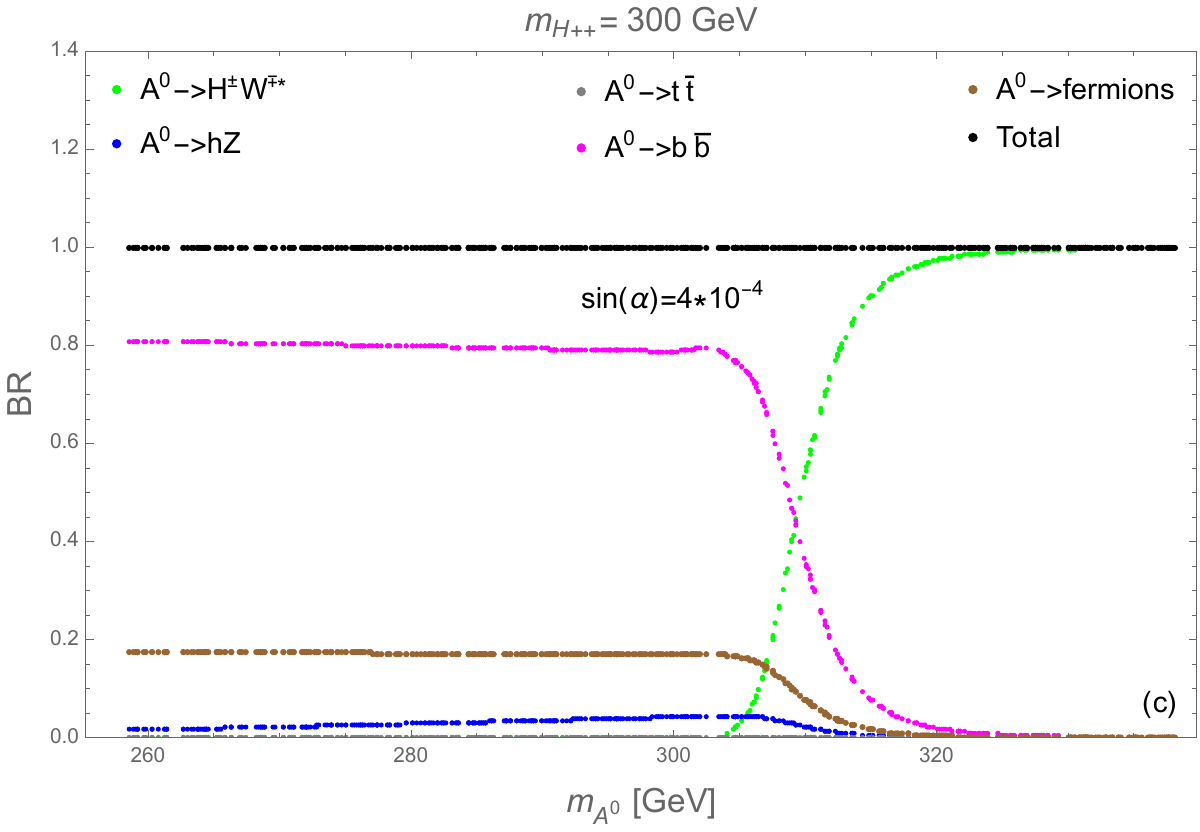}}
    \end{center}
\caption{\label{fig:BRs-2} Branching ratios for different Higgs scalars of the model, for \mHpp $=300$~GeV, $v_t=0.1$~GeV and \sina $=1.2 \times 10^{-3}$ in (a), (b), and \sina $=4 \times 10^{-4}$ in (c).}
\end{figure}

A complex decay pattern due to a high sensitivity to $\sin \alpha$ of some decay widths is observed for most of the scalars branching ratios (BRs). 
This is one of the novelties of the study.
   Some observations about the BRs behaviour are presented below:  

\begin{itemize}
\item {\bf {\boldmath \Hpp} scalar boson}: The \Hpp$\to W^\pm W^\pm$ decay being on-shell, it dominates the BRs for most of the $H^\pm$ mass range, as we see in~\cref{fig:BRs}(a). There are also off-shell decays, notably \Hpp $\rightarrow H^\pm W^{\pm *}$, where the off-shell $W^\pm$ is due to the small mass splitting $|m_{H^{\pm 
\pm}} - m_{H^\pm}| \leq 20$~GeV, to cope with electroweak precision data .  We show in the same figure separate contributions for the leptonic or hadronic $W^{\pm *}$ decays; between them, an expected factor 2 difference is observed. Let us note that the \Hpp BRs do not have any \sina dependence, the corresponding widths being controlled exclusively by the gauge couplings and the magnitude of $v_t$.

\item {\bf \boldmath{\Hp} scalar boson}:  In this case we see an involved dependence on \sina. All the \Hp BRs possibilities are depicted in~\cref{fig:BRs}(b) and \cref{fig:BRs-2}(a), where different \Hp mass regions are dominated by different BRs. The on-shell channels, $tb$, $WZ$ and $h^0W^\pm$ compete in the central region of \Hp masses, as soon as they become kinematically available. The extremes are then taken by $H^\pm \rightarrow H^0/A^0 W^{\pm *}$ (in the lower part of the \Hp mass spectrum) or   $H^\pm \rightarrow H^{\pm \pm} W^{\pm *}$ (in the higher part of the \Hp mass spectrum) that lead to further cascade decays.  This order remains more or less the same with increasing \mHpp, until \sina rises well above the value that minimizes the $H^\pm W^\mp h^0$ coupling, after which $H^\pm \rightarrow h^0W^\pm$ dominates for most of the mass range, as shown in~\cref{fig:BRs-2} (a).

\item {\bf \boldmath${H^0}$ scalar boson}: \cref{fig:BRs}(c) illustrates the case where the on-shell $ZZ$ decay dominates over most of the $H^0$ mass spectrum, until $H^0 \rightarrow H^\pm W^{\mp *}$  with an off-shell $W^\pm$ kicks in for heavier $H^0$.   
%This last BR is also the most stable one across different \sina values, since it has no \sina dependence. 
In fact the chosen value of \sina shuts off the coupling controlling the $W^\pm W^\mp$ channels. A value twice bigger would have enhanced the latter channel and suppressed the $ZZ$ one. However for such larger \sina the $h^0h^0$ channel becomes dominant, the two others suffering partial cancellations in the couplings. This effect is illustrated
in~\cref{fig:BRs-2} (b).

\item  {\bf \boldmath${A^0}$ scalar boson}:  In contrast to $H^0$, the CP-odd $A^0$ has no tree-level decays to $ZZ$ or $W^\pm W^\mp$. For the \sina value and mass range selected in \cref{fig:BRs}(d), the dominant decay is to $h^0Z$ for most of the $A^0$ masses---this width being the only \sina-dependent one. It scales as $\sin^2 \alpha$ for the chosen range of \sina$ > v_t/v_d$ which explains its large dominance over the $b\bar b$ channel, the latter having both $(v_t/v_d)^2$ and $(m_b/m_{A^0})^2$ suppressions. The other part of the spectrum is dominated by the cascade decays of $H^\pm W^{\mp *}$. \Cref{fig:BRs-2}(c) illustrates the opposite configuration where the reduced \sina value tends to shut off the coupling that controls $A^0 \to h^0Z$, enhancing tremendously the $b\bar b$ channel. The $t \bar t$ channel would open for higher mass scales.

\end{itemize}
All the above features can be understood from the structure of the couplings and the mass dependence, keeping in mind that 
\sina and $v_t$ are two independent parameters, as explained in detail in ~\cite{ducu}.

\FloatBarrier

\section{The intermediate states}
\label{sec:intermediate}

For the highest cross-sections observed according to section~\cref{sec:production_decay}, 
the decay BRs of each scalar's pair or associated production, be it charged, neutral or mixed,  combine in complex ways.   Instead of listing the intermediate \footnote{the SM particles that can decay further to leptons, jets, photons or neutrinos. These can be: SM Higgs boson, top quarks, W and Z bosons} or final  states, we propose a list of criteria with which the intermediate and then final states can easily be derived. Firstly,  only two mass hierarchies  are possible: 
   % \begin{align}
      $m_{H^{\pm\pm}} \geq m_{H^{\pm}} \geq m_{H^0} \simeq m_{A^0}$ or 
       $m_{H^{\pm\pm}} \leq m_{H^{\pm}} \leq m_{H^0} \simeq m_{A^0}$, 
       %\label{eq:h}
  %  \end{align}
thus only two possible configurations of decay chains of one state to another are possible. Then, depending on the scalar mass scale, direct decays to SM states can become available. Thirdly, the decays of most of the BSM scalars have a certain degree of sensitivity to \sina.

Based on these considerations,  one is led to take into account three decay chains:  long-chain decays (LC) where three scalar states are involved in the cascade, intermediate-chain decays (IC) where only two are involved, and finally  direct-chain decays (DC) where the scalar state decays directly to SM states, that is $W$'s, $Z$'s, tops, bottoms and $h^0$'s.
Together with the production channels enumerated in~\cref{sec:production_decay} and the other two criteria mentioned before, 
%which refer to the mass scale of the scalar bosons and \sina dependence of the BRs, 
one can derive all the intermediate states, and then all the final states of interest for an ATLAS/CMS analysis. Details about the procedure can be found in~\cite{ducu}. It should be mentioned that most of the intermediate states are equally promising, due to the fact that given the \sina sensitivity, varying it slightly promotes to dominant virtually any of the decay patterns.

\section{Illustrative benchmark points for a potential experimental search at the LHC}
\label{sec:analysis}

The experimental search potential of LHC is assessed by examining the \Hpp \Hmm, \Hpp \Hm and $A^0 H^0$ production modes, which have the highest cross-sections on a wide range of \mHpp masses. We focus on multi-lepton experimental signatures, which have a smaller SM background. In particular, we study the states of two leptons of the same electric charge (SC), three leptons, or four leptons, with jets and missing transverse energy.  

Signal samples of 50.000 events were generated in MG5\_aMC\_v3.5.3~\cite{alwall} and then showered and hadronized with  Pythia\_v8~\cite{bierlich}. Madspin~\cite{artoisenet} was also used to ensure that both spin-correlations and off-shell effects are kept, and then Delphes~\cite{delphes} framework to mimic the interaction of hadronized matter with the the ATLAS detector. The obtained samples were analyzed using the SimpleAnalysis framework~\cite{ATLAS2022}, with which yields at various selection stages, region acceptance and statistical significances of the signal were estimated. 

Further we will present only some general conclusions of prospected searches performed for some of the production channels stated in section~\cref{sec:production_decay}, 
for three benchmark points: \mHpp = 220, 300 and 400 GeV. The analyses are based on a recent ATLAS collaboration study~\cite{ATLAS2021},
the details being in depth described in~\cite{ducu}:

\begin{itemize}
\item \Hpp \Hmm production mode: a 100\% decay BR into $W^\pm W^\pm$ is considered.
As expected, a drastic decrease in the number of events (and acceptance) is observed when multi leptons are required. Computing the signal significance $Z$ using the number of background events from Ref~\cite{ATLAS2021}, the 220~GeV mass point can already be excluded by the present LHC energies and luminosities, whereas for HL-LHC luminosities all 3 mass points can be excluded.  

\item \Hpp \Hm production mode: we considered the $H^\pm \rightarrow W^\pm Z$ and $H^{\pm \pm} \rightarrow W^\pm W^\pm$ decay modes. The 220 GeV mass point can already be excluded, whereas for 14 TeV and 300 fb$^{-1}$, mass points up to 400 GeV could be
excluded and points up to 300 GeV discovered (if the model is
true) with no changes in analysis.

\item $H^0 A^0$ production mode, which is looked at for the first time: we observe that for $A^0 \rightarrow H^\pm W$, $H^\pm \rightarrow WZ$ and $H^0 \rightarrow ZZ$ decay modes, the statistics are reasonable only for lower mass
points, thanks to the higher cross-sections. In addition, as a high amount of SM background is expected in the 0, 1 and 2 leptons channels, we concluded that the the most sensitive channel could be the one with 3 leptons. However, when considering the $H^0 \rightarrow ZZ$, $A^0 \rightarrow H^\pm W$, $H^\pm \rightarrow H^{\pm \pm} W$, $H^{\pm \pm} \rightarrow WW$ decay modes, the 2$\ell$ SC and 3 leptons channels appear promising.
\end{itemize}

\section{Conclusions}

In this work we investigated the scalar sector of the Type 2 Seesaw Model for which the decay patterns are complex. We pinpointed a high sensitivity to the mixing angle in the CP-even sector which was not previously emphasized, suggesting the need for a global experimental search strategy. Based on the identified promising channels, we studied the ATLAS-like search potential for the charged and neutral sectors with 2SC-, 3- and 4-lepton selections. 
%For the charged sector, some mass points can already be excluded,  as expected from the experimental studies shown in \cite{ATLAS:2021pairbosons,Marseille:2024ddj}. As for the neutral sector, which has not been studied before, 3-lepton channels seem to have some clear advantages. 

The aim of this paper was thus to achieve an exhaustive phenomenological study to identify the most promising search channels, which will hint at interesting
directions for the future LHC and HL-LHC experimental searches. 

%These estimates give possible directions to be followed for the search of scalar particles specific to this model. 

%....

%\bibliographystyle{JHEP}
%\bibliography{references}

\acknowledgments 
This work received support from the French government under the France 2030 investment plan, as part of the Excellence Initiative of Aix Marseille University - A*MIDEX (AMX-19-IET-008 - IPhU), and  support from IFIN-HH under the Contract ATLAS CERN-RO with the Romanian MCID / IFA. GM has received partial support from the European Union’s Horizon 2020 research and innovation programme under the Marie Skłodowska-Curie grant agreements No 860881-HIDDeN and No
101086085–ASYMMETRY.

\end{document}